\documentclass[twocolumn,showpacs,preprintnumbers,amsmath,amssymb,aps,dvips]{revtex4}
\usepackage{amsmath}
\usepackage[dvips]{graphicx}
\usepackage{dcolumn}
\newcolumntype{d}[1]{D{.}{.}{#1}}
\usepackage{bigdelim}
\usepackage{bm}
\usepackage{array}
\usepackage{longtable}
\usepackage{lscape}
\usepackage{float}
\usepackage{rotating}
\usepackage{multirow}
\usepackage{setspace}

\makeatletter

\newcommand{\Rmnum}[1]{\expandafter\@slowromancap\romannumeral #1@}
\makeatother

\begin{document}

\preprint{APS/123-QED}

\title{Space Reduction in Matrix Product State}
\author{Lihua Wang$^{1}$}
\email{wanglihua94@tsinghua.org.cn}
\author{Kwang S. Kim$^{1}$}
\email{kim@unist.ac.kr}
\affiliation{$^1$
Dept. of Chemistry, School of Natural Science, Center for Superfunctional Materials, Ulsan National Institute of Science and Technology (UNIST), Ulsan 44919, Republic of Korea}

\date{\today}

\begin{abstract} 
We reconstruct a matrix product state (MPS) in reduced spaces using density matrix. This scheme applies to a MPS built on a blocked quantum lattice. Each block contains $N$ physical sites that have a local space of rank $R$. The simulation in the original spaces of rank $R^N$ is used to construct density matrices for every block. They are diagonalized and only the eigenvectors corresponding to significant diagonal elements are used to transform the original spaces to smaller ones and to reconstruct the MPS in those smaller spaces accordingly. Simulations in the reduced spaces are used to reliably extrapolate the result in unreduced spaces. Moreover, to obtain a required accuracy, the ratio of the reduced space rank over the original decreases quickly with $N$. The reduced space has a saturated rank to obtain a demanded accuracy when $N\rightarrow \infty$.    
\end{abstract}

\maketitle

The use of density matrix is crucial in simulating quantum lattice model in that it guides reduction of Hilbert space/subspace which exponentially increases with respect to the system/subsystem size. Given a bipartite structure $A$ and $B$ of a system, the system wave function $\mid \Psi\rangle$ is an entangled quantity composed of basis vectors from subspaces $\{\mid \phi_i\rangle\}$ for $A$ and $\{\mid \psi_i\rangle\}$ for $B$, 
\begin{equation}
	\label{eq:systemwavefunction}
	\mid \Psi\rangle=\sum_{i,j}{X_{ij}\mid \phi_{i}\rangle\mid \psi_{j}\rangle}
\end{equation}
where $X$ is a tensor entangling $A$ and $B$. The density operator for a subspace, say $\{\mid \phi_i\rangle\}$, is defined as $\hat{\rho}\equiv Tr_B\mid \Psi\rangle\langle \Psi\mid$, where $Tr_B$ means that the degree of freedom in subsystem $B$ is traced out. Its matrix representation is 
\begin{align}
\label{eq:densitymatrix}
\left(\rho_{ij}\right)\equiv \langle \phi_i\mid \hat{\rho}\mid \phi_j\rangle=X_{ik}X^*_{jk}
\end{align}
. Hereafter, Einstein summation convention is implied for repeating indices in a formula. Note that normalization of $\mid \Psi\rangle$ requires $tr\rho=1$. 

The system wave function can be reconstructed as $\mid \Psi^{\prime}\rangle$ using a reduced space $\{\mid\theta_i\rangle\}$ consisted of $M$ basis vectors for $A$ along with the unaltered space $\{\mid \psi_i \rangle\}$ for $B$. The density matrix built in $\{\mid \phi_i\rangle\}$ is used to make the residual vector $\mid R\rangle\equiv \mid \Psi\rangle-\mid \Psi^{\prime}\rangle$ have a minimum norm\cite{Schollwoeck2005,White1992,White1993}. Explicitly, the density matrix is diagonalized and only $M$ eigenvectors $\{v_i;i=1,\cdots ,M\}$ need to be retained. They correspond to the most significant $M$ diagonal elements $\{\eta_i\}$. New basis vector is constructed as
\begin{align}
\label{eq:basis_transformation}
\mid \theta_i\rangle=v_i\left(k\right)\mid \phi_k\rangle
\end{align}
where $v_i\left(k\right)$ is the $k^{th}$ element of the eigenvector $v_i$. And, $\mid \Psi^{\prime}\rangle$ is constructed as
\begin{align}
\label{eq:reduced_function}
\mid \Psi^{\prime}\rangle=Y_{ij}\mid \theta_{i}\rangle\mid \psi_{j}\rangle
\end{align}
where
\begin{align}
\label{eq:function_transformation}
Y_{ij}=v_i\left(k\right)X_{kj}
\end{align}
. If the formal reduction is unitary (zero reduction), substituting equations \eqref{eq:basis_transformation} and \eqref{eq:function_transformation} into equation \eqref{eq:reduced_function} restores the wave function in equation \eqref{eq:systemwavefunction}. When the truncation of space of $A$ takes place, i.e, the eigenvector matrix kept is rectangular, hence no longer unitary, we have
\begin{equation}
\label{eq:restore_function}
\mid \Psi^{\prime}\rangle = X_{kj}v_i(k)v_i(l)\mid \phi_{l}\rangle\mid \psi_{j}\rangle = \mid \Psi\rangle-\mid R\rangle
\end{equation} 
where
\begin{equation}
	\label{eq:residual}
	\mid R\rangle=X_{kj}\Delta_{kl}\mid \phi_{l}\rangle\mid \psi_{j}\rangle
\end{equation}
. Here, $\Delta_{ij}=\sum_{k=M+1}^{n}v_i\left(k\right)v_j\left(k\right)$. It is straightforward to show $||\mid R\rangle||^2=\sum_{i=M+1}^{n}{\eta_i}$.

Following the line of local space reduction, the density matrix renormalization group (DMRG)\cite{White1992,White1993} uses the density matrix to keep a fixed amount of transformed basis vectors for an enlarged part of system. Meanwhile, the density matrix embedded theory (DMET)\cite{Knizia2012} provides an alternative to the dynamic mean field theory (DMFT)\cite{Metzner1989}, using the density matrix to improve the impurity state of a fragment embedded in a background.   

We implement the density matrix in a different way where it is used to reduce spaces in a matrix product state (MPS)\cite{Fannes1992,Oestlund1995,Verstraete2004,Chung2006,Chung2007,Chung2009,Wang2012,Garcia2007,Crosswhite2008,McCulloch2008a}. Dividing a quantum lattice into $L$ blocks each of which contains $N$ physical sites, a MPS is built as 
\begin{align}
\label{eq:mps}
\mid \Psi\rangle=\sum_{\cdots r^i\cdots r^L}{tr\left(\cdots \xi_{r^{i-1}}^{i-1}\cdot \xi_{r^i}^i\cdots\right)\cdots\mid \phi_{r^{i-1}}^{i-1}\rangle\mid \phi_{r^{i}}^{i}\rangle\cdots} 
\end{align} 
where each MPS tensor $\xi^i$ is associated with a block, say, the $i^{th}$ block. It has three indices. The first index $r^i$ runs from $1$ to $Q\equiv R^N$ because it is associated with $N$ physical sites in a block, each of which has a space ranked $R$. The other two are the bond indices participating matrix product and are not explicitly shown in equation \eqref{eq:mps}. They run from $1$ to $P$ where $P$ is the MPS rank characterizing the entanglement in the wave function\cite{Schollwoeck2005}, the larger of which gives the more precise representation of entanglement present in the system. 

The computational burden of variational optimization of each MPS tensor is determined by both $P$ and $Q$. One needs a tractable strategy to balance between choices of $P$ and $Q$. Choosing blocks that contain more physical sites has the following benefits. First, there are fewer tensors to solve. Second, it uses smaller $P$ to achieve a comparable precision. In the extreme case when a block contains the whole system, one just needs a MPS tensor of rank $1$ to precisely represent the system wave function. However, as $Q$ exponentially increases with $N$ to exclude possibility of building MPS on a block containing the whole system, one still needs to solve multiple MPS tensors, while the same computational resource only allows smaller $P$ when $N$ is larger. 

We propose a scheme to overcome this difficulty when building MPS on a blocked quantum lattice. A MPS ranked $P_1$ in the original spaces, $\mid \Psi\rangle_{\perp P= P_1}$, is used to construct density matrix for each block, say the $i^{th}$ block, as follows
\begin{widetext}
	\begin{align}
	\label{eq:construction}
	\rho_{ab}^{i}=tr\left[\cdots\left(\xi^{i-1}_{r^{i-1},\alpha_1\gamma_1}\xi^{i-1}_{r^{i-1},\alpha_2\gamma_2}\right)\left(\xi^i_{a,\gamma_1\eta_1}\xi^i_{b,\gamma_2\eta_2}\right)\left(\xi^{i+1}_{r^{i+1},\eta_1\theta_1}\xi^{i+1}_{r^{i+1},\eta_2\theta_2}\right)\cdots\right]
	\end{align}
\end{widetext}
. In $\xi^i_{a,\gamma_1\eta_1}$, $a$ denotes the space index and the bond indices $\gamma_1$ and $\eta_1$ are explicitly shown here. Density matrices will be constructed for all $L$ tensors (blocks) and are diagonalized simultaneously. For each block, only eigenvectors corresponding to the most significant $M$ diagonal elements are used to transform the space $\{\mid \phi^i_j\rangle\}$ to the smaller one $\{\mid \theta^i_j \rangle\}$ according to equation \eqref{eq:basis_transformation}, where the superscript refers to the block's label.

Note that MPS in the reduced space set $\{H^{i,\prime}\equiv\{\theta^i_j\}\}$ for $P\le P_1$, $\mid \Psi^{\prime}\rangle_{\perp P\le P_1}$, can be reconstructed from the existing MPS $\mid \Psi\rangle_{\perp P\le P_1}$ as 
\begin{align}
\label{eq:newmps}
\mid \Psi^{\prime}\rangle=\sum_{\cdots s^i\cdots s^L}{tr\left(\cdots \kappa_{s^{i-1}}^{i-1}\cdot \kappa_{s^i}^i\cdots\right)\cdots\mid \theta_{s^{i-1}}^{i-1}\rangle\mid \theta_{s^{i}}^{i}\rangle\cdots} 
\end{align}
where 
\begin{align}
\label{eq:newmpstensor}
\kappa^i_{a}=v_a\left(b\right)\xi^i_{b}
\end{align}
. $\mid \Psi^{\prime}\rangle_{\perp P> P_1}$ is then variationally determined using $\mid \Psi^{\prime}\rangle_{\perp P= P_1}+\mid \delta\rangle$, where $\mid \delta\rangle$ is a small perturbation, as a trial wave function\cite{Wang2018}. Or, the MPS in reduced spaces can be variationally determined for all the range of $P$. In both ways, since $\{H^{i,\prime}\}$ has a smaller rank $Q^{\prime}<Q$, the same computational resource now allows larger $P$, yielding better accuracy in turn.
\begin{figure}
	\begin{center}
		$\begin{array}{cc}
		\mbox{(a)}&\\
		& \includegraphics[width=16.pc]{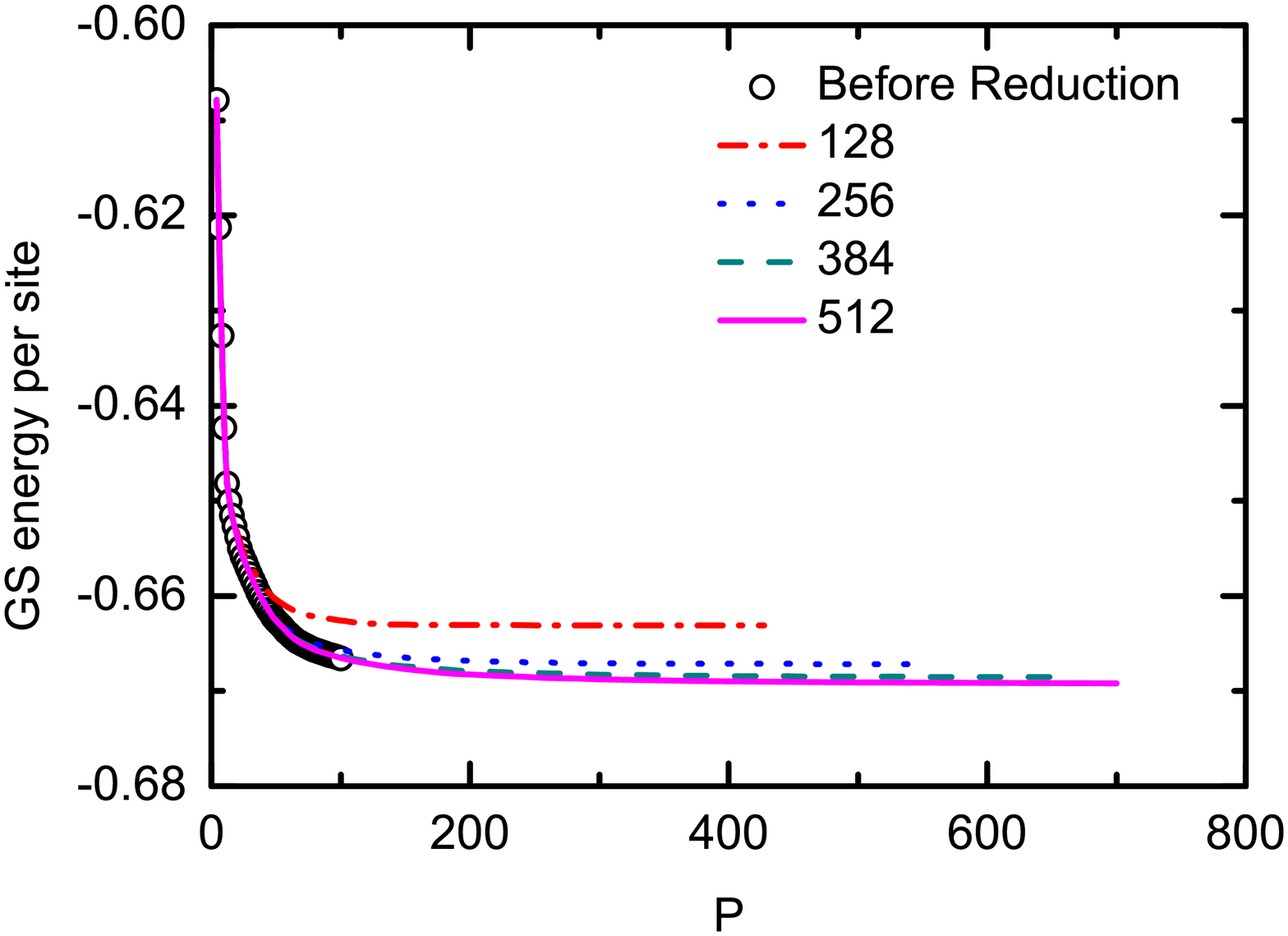}\\
		\mbox{(b)}&\\	
		& \includegraphics[width=16.pc]{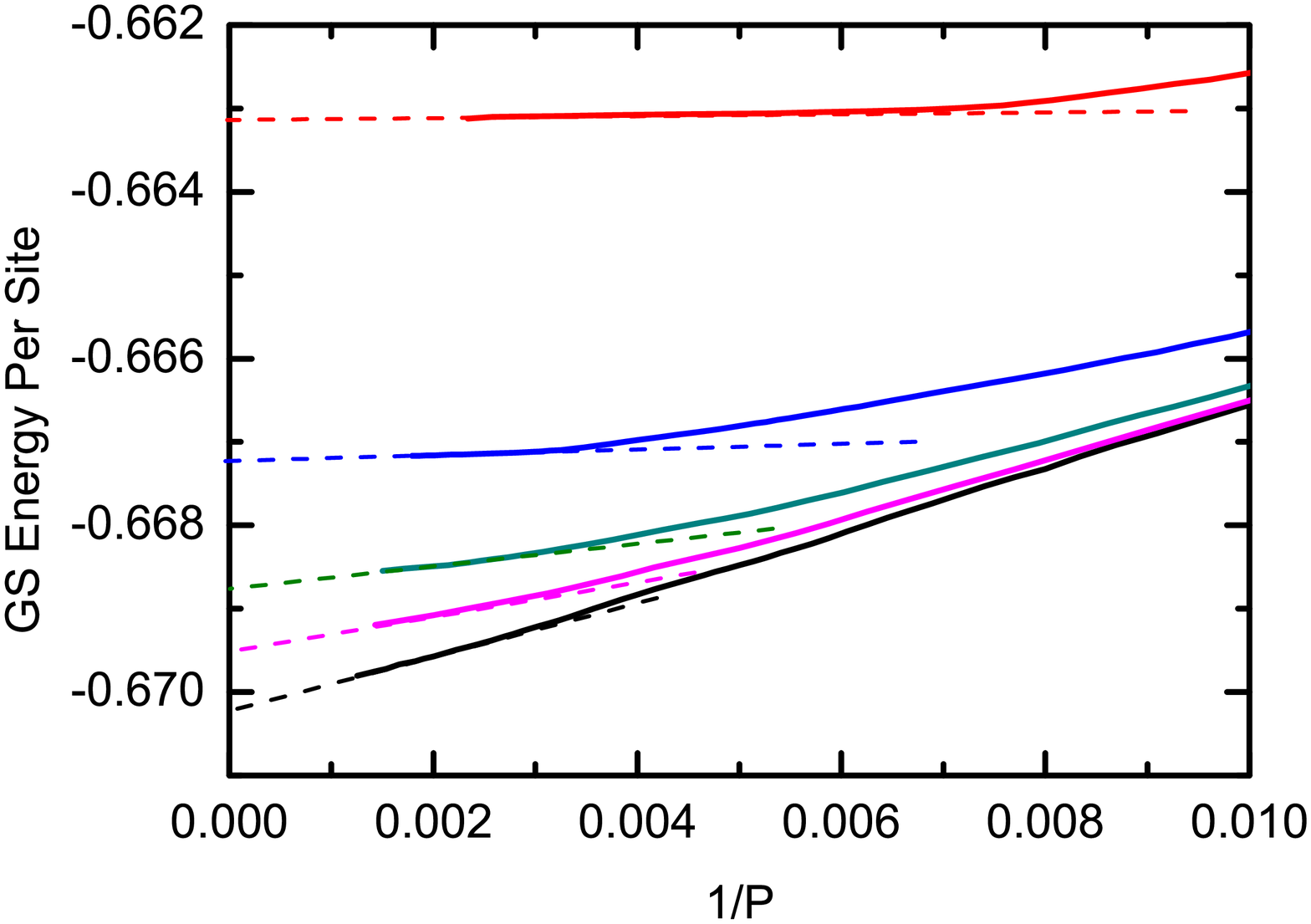}\\
		\mbox{(c)}&\\	
		& \includegraphics[width=16.pc]{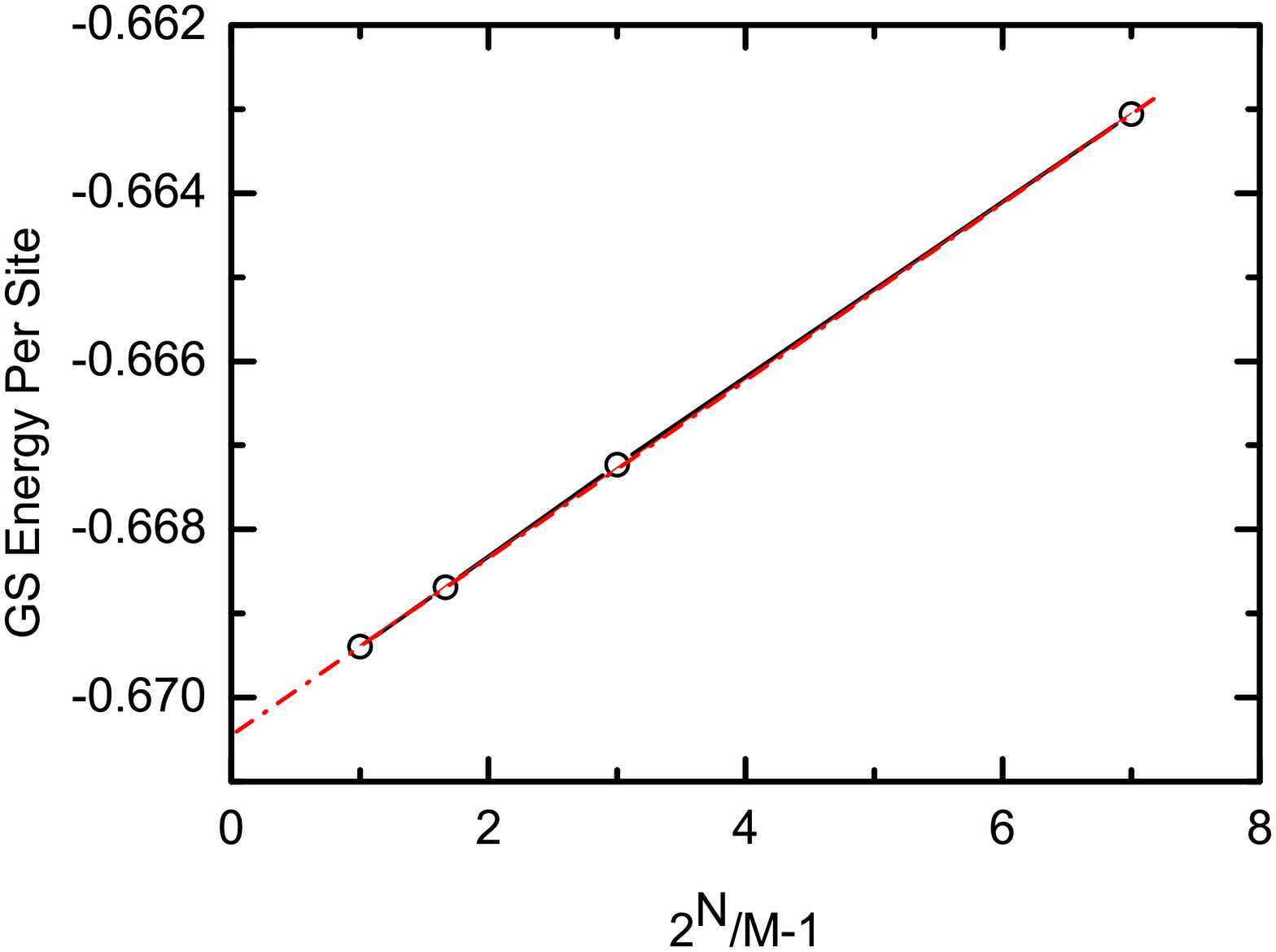}		
		\end{array}$
		\caption{\label{fig:energy}Effect of space reduction in MPS for $N=10$. (a) GS energy per site with respect to MPS rank $P$. Open circles denote the solution before reduction. At $P_1=100$, such a solution is used to reduce the MPS space rank to $128$, $256$, $384$ and $512$, yielding new solutions shown as dot-dashed, dotted, dashed and solid curves, respectively. (b) GS energy versus $1/P$. Tangents of convergence yield the extrapolated energies for various space ranks, $128$, $256$, $384$, $512$ and $1024$ (unreduced) from top to bottom. (c) Extrapolation of the energy in unreduced spaces using those obtained in reduced spaces.}
	\end{center}
\end{figure}   
\begin{figure}
	\begin{center}
		$\begin{array}{c}
		\includegraphics[width=16.pc]{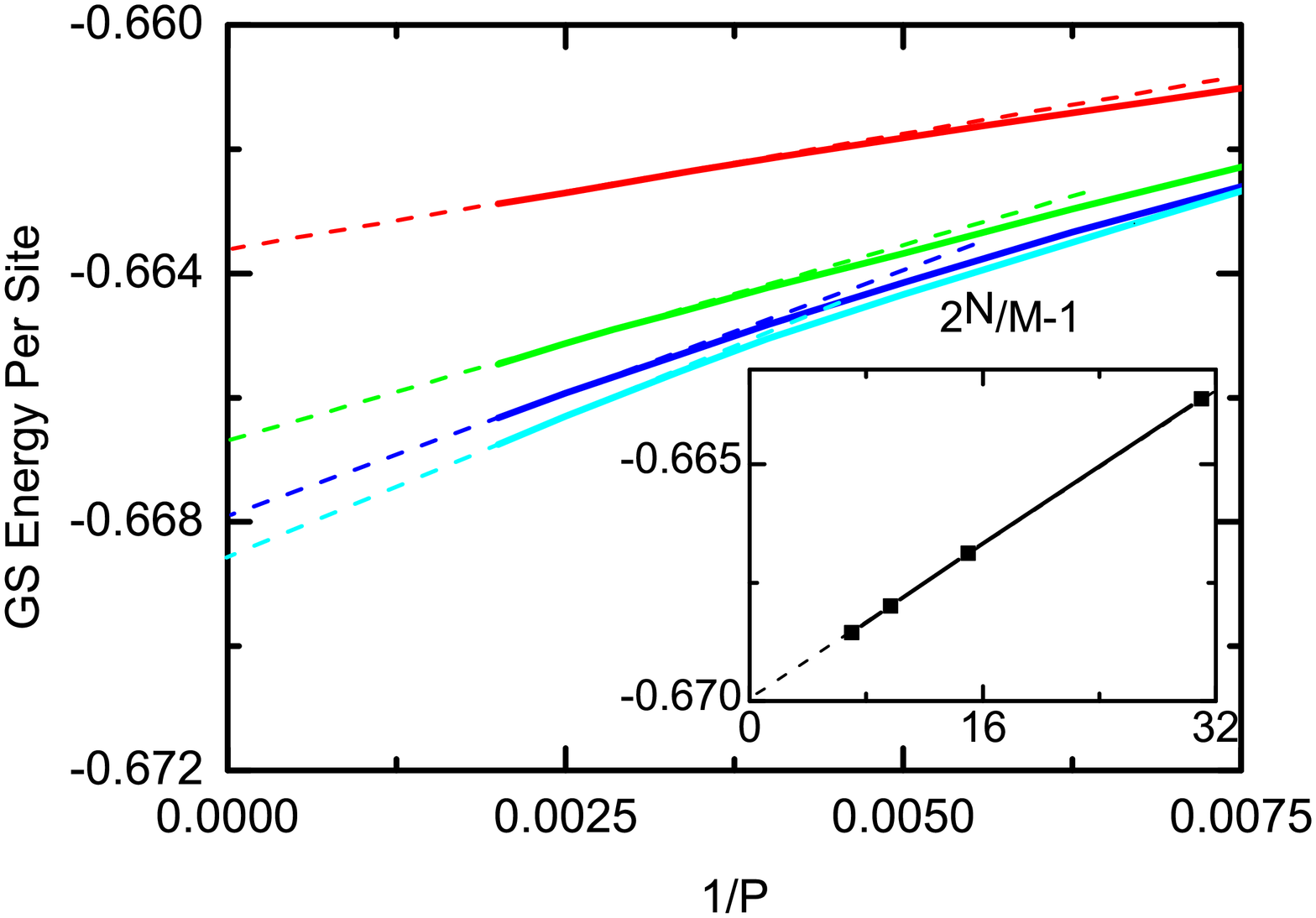}\\
		\end{array}$
		\caption{\label{fig:energy1}Use of space reduction in MPS for $N=14$. Tangents of the convergence of GS energy per site versus $1/P$ yield extrapolated energies for various space ranks, $512$, $1024$, $1536$ and $2048$ from top to bottom. The inset extrapolates GS energy per site in the unreduced space using those extrapolations obtained in reduced spaces.}
	\end{center}
\end{figure}  
  
We apply this approach to solve the antiferromagnetic spin $\frac{1}{2}$ Heisenberg model on the infinity by $N$ square lattices\cite{Wang2018}. The Hamiltonian is
\begin{equation}
\label{eq:hamiltonian}
H=J\sum_{\langle \left(i,j\right),\left(i',j'\right)\rangle}{{\vec S}_{\left(i,j\right)}\cdot {\vec S}_{\left(i',j'\right)}} 
\end{equation} 
where ${\vec S}_{\left(i,j\right)}$ is the spin vector operator on the ${\left(i,j\right)}^{\text{th}}$ lattice site with $\it{i}$ running from $-\infty$ to $\infty$ in the longer dimension (LD), while $\it{j}$ runs from $1$ to $N$ in the shorter dimension (SD). $\langle\rangle$ sums over the nearest neighboring sites. $J$ is the spin-spin coupling integral and is normalized to $1$ hereafter. The periodic boundary condition is assumed in both LD and SD. 
\begin{figure}
	\begin{center}
		$\begin{array}{cc}			
	\mbox{(a)}&\\
& \includegraphics[width=16.pc]{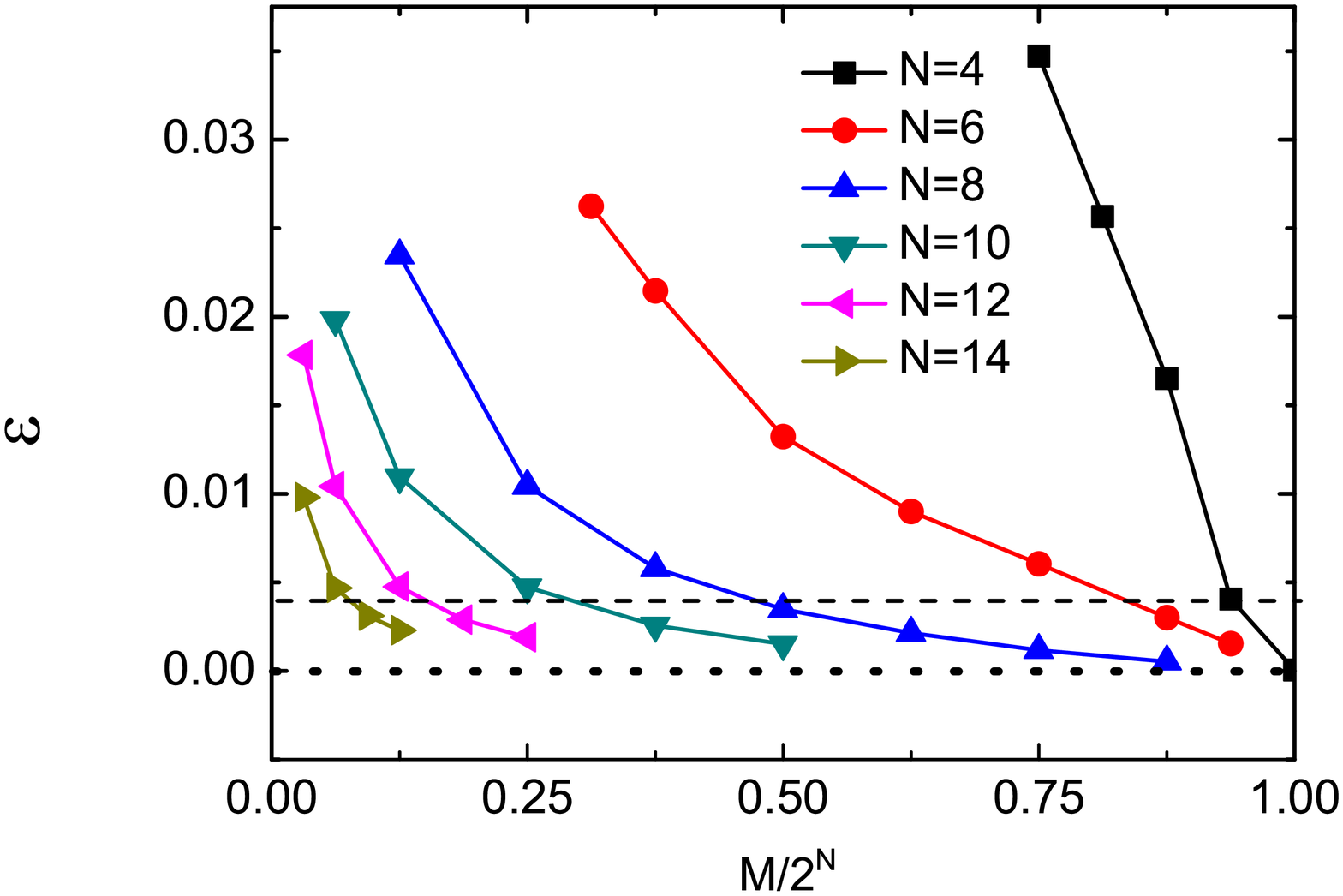}\\
	\mbox{(b)}&\\	
& \includegraphics[width=16.pc]{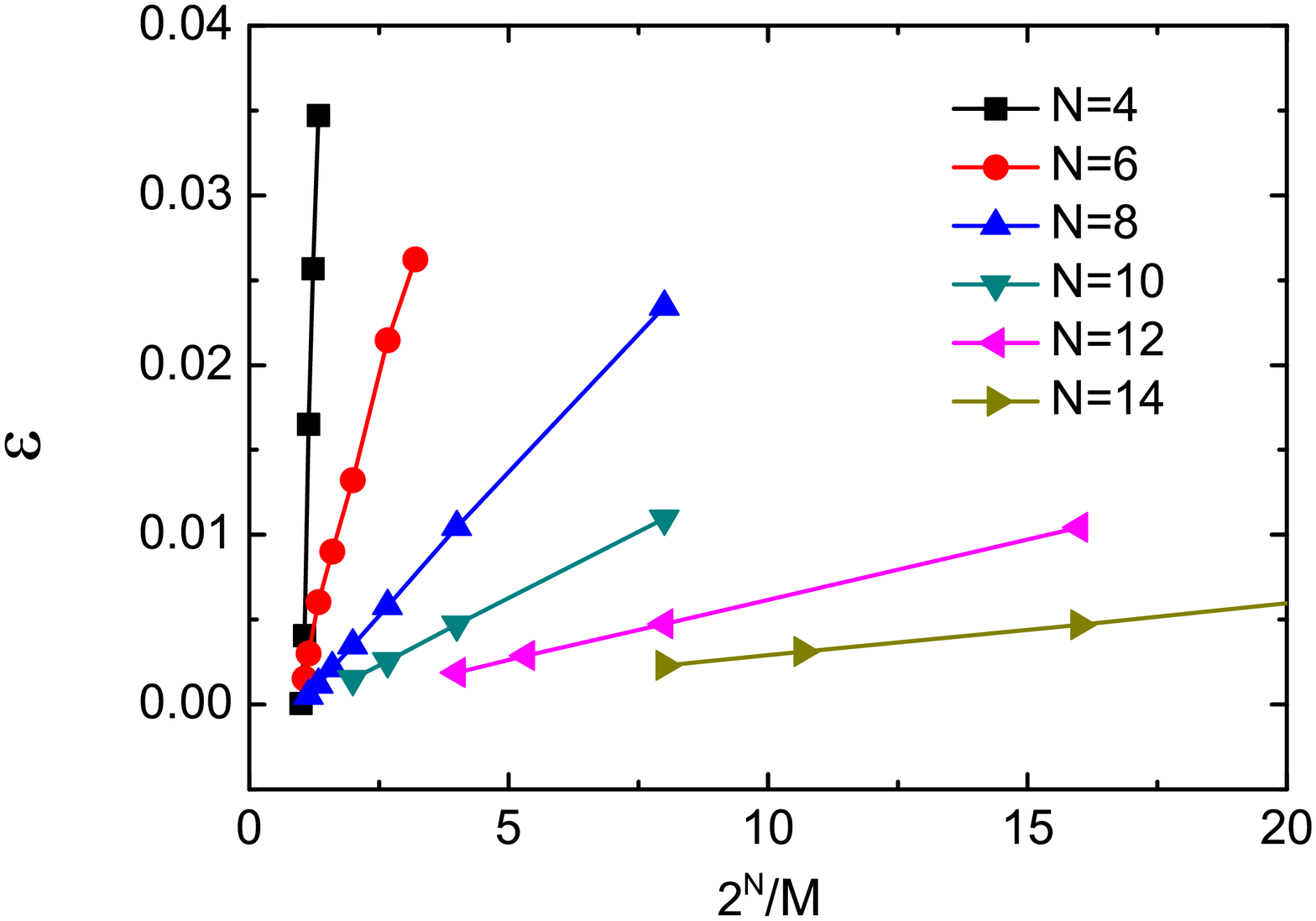}\\
		\end{array}$
		\caption{\label{fig:spaceerror} Relative error versus (a) $M/2^N$ (reduction ratio) and (b) $2^N/M$ (inverse ratio). In (a), The dotted line gives a reference of zero error, while the dashed line intercepts each curve to give the reduction ratio at a certain accuracy. (b) Energies obtained in reduced spaces for various infinite by $N$ lattices all approach linearly to those in unreduced space. Fig.\ref{fig:energy}(c) shows such an example for $N=10$. Fig.\ref{fig:energy1} shows another example for $N=14$.}
	\end{center}
\end{figure} 

Previously, the infinite quasi-1D entanglement perturbation theory (iqEPT)\cite{Wang2015} was extended in\cite{Wang2018}. It pinpointed a quantum dimension transition caused by the spontaneous spin rotational symmetry breaking and obtained GS energy with a relative error less than $2\times 10^{-3}$ for $N$ up to $12$. In iqEPT, the model is treated as if in one dimension (1D). $N$ physical sites in SD are blocked into one effective site. A MPS is built on those blocks and the entanglement in it saturates when $N$ is large. Meanwhile, the Hamiltonian is expressed as a matrix production operator (MPO)\cite{Crosswhite2008,Chung2007,Chung2009,Wang2012,Wang2015,Verstraete2004a,Pirvu2010,Schollwoeck2005}. The linking complexity between the building units of MPO is reduced to a linear dependence on $N$. The reduction of block's space of rank of $2^N$ is crucial in handling even larger lattices or different complex scenarios by iqEPT, because it is the only remaining factor that gains exponential complexity with respect to the lattice size. 

We use the extended iqEPT\cite{Wang2018} as the solver for simulations both before and after space reduction. The effect of space reduction in MPS is shown with the example of $N=10$ in Fig.\ref{fig:energy}. In (a), simulated data in the original space of rank of $2^{10}$ are shown as open circles. At $P_1=100$, the solution is used to reduce the space rank to $128$, $256$, $384$ and $512$ to yield solutions in dot-dashed, dotted, dashed, and solid curves, respectively. Except the reduced rank $128$, simulations for other reductions reproduce the solution before reduction when $P\le P_1$. The closing gaps between flattening curves are confirmed in (b), where the energies versus $1/P$ are plotted for space ranks $128$, $256$, $384$ and $512$ from top to bottom. The simulation in original space is also carried on after $P_1$, shown as the bottom curve in the same plot. All curves show convergency. The extrapolation by tangents of those converging curves yields energies in spaces of both various reduced sizes and the original size. Those in the reduced spaces are used to extrapolate the energy in the unreduced space, as shown in (c). There, the linear fit yields $0.6704$, agreeing well with $0.67022$ by extrapolation using the data obtained before space reduction in (b). Note that this scheme which extrapolates the result in the original space with the data obtained in reduces spaces, is much more computationally efficient so as to allow simulation at larger $P$ values. We run simulations for $N=14$ in various reduced spaces of ranks $512$, $1024$, $1536$ and $2048$ up to $P=500$, shown from top to bottom in Fig. \ref{fig:energy1}. The lowest energy without extrapolation is $-0.66676$ at $P=500$ in the reduced space of size $2048$, lower than the previously reported value of $-0.66636$ at $P=350$ that is the largest $P$ value handleable in the unreduced spaces. Meanwhile, the inset extrapolates to $-0.66998$ with the difference of $5 \times 10^{-5}$ from $-0.66993$ which was obtained by the interpolation in Fig. 6(c) of \cite{Wang2018}.   
  
Fig.\ref{fig:spaceerror}(a) shows the result of $\epsilon$ versus $M/2^N$ for various $M^{\prime}s$ and $N^{\prime}s$, where $2^N$ is the original space rank; $\epsilon$ is the relative error between the energies obtained before and after reduction. It is seen that only $1/8$ of the original space size $2^{12}$ is needed for $N=12$, to achieve a relative error of $4.1\times 10^{-3}$. In comparison, the same accuracy for $N=4$ is obtained with $15$ out of $2^4$ basis vectors. For $N=2$, no reduction will achieve good accuracy. Fig.\ref{fig:spaceerror}(b) shows the dependence of relative error on $2^N/M$. Larger lattices ($N\ge 8$) show a linear dependence, which is a reconfirmation for the reliability of extrapolating results using simulation in reduced spaces. Fig.\ref{fig:energy}(b) and (c) illustrate such an example for a lattice of $N=10$. Fig.\ref{fig:energy1} shows another example for $N=14$.
 
\begin{figure}
	\begin{center}
		$\begin{array}{ccc}			
		&\includegraphics[width=16.pc]{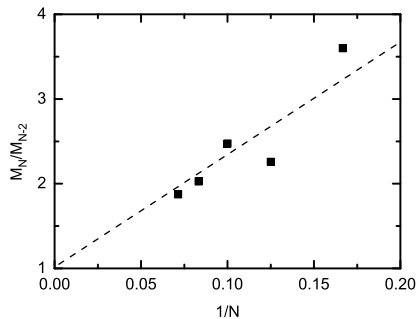}& \\
		\end{array}$
		\caption{\label{fig:fixedpoint} $M_N/M_{N-2}$ (ratio of numbers of basis vectors kept for $N$ and $N-2$, respectively) versus $1/N$. The linear fit overlaps with the guiding dashed line to $1$ when $N\rightarrow \infty$. The quantum dimension transition\cite{Wang2018} is responsible for the kink of the guiding line when $N=8$.}
	\end{center}
\end{figure}

At last, we plot in Fig.\ref{fig:fixedpoint} $M_N/M_{N-2}$ (ratio of numbers of basis vectors kept to achieve the same accuracy for $N$ and $N-2$, respectively) versus $1/N$. It shows that this ratio tends to approach $1$ when $N\rightarrow \infty$. Therefore, an $\infty \times \infty$ lattice can be treated as if in 1D, when looked from any of its two dimensions. Possibly, only a fixed number of basis vectors after transformation by density matrix is needed for a demanded accuracy.

In conclusion, we show that the space reduction in MPS can be used to extrapolate reliable results using the data obtained in the reduced space which is easier to deal with. Moreover, it changes the remaining exponentially increasing factor in iqEPT\cite{Wang2018} to a fixed factor. Since there is no more exponentially increasing factor, it is a promising method for strong correlations in low dimensions. Note that the space reduction in MPS shown in this study can be readily extended to any form of MPS or tensor network state (TNS)\cite{Nishino2000,Nishio2004} based methods such as the projected entangled pair state (PEPS)\cite{Verstraete2004,Verstraete2004a,Verstraete2004b,Jordan2008}, whenever they are built on a blocked quantum system.     
  
\hspace{11cm}

\begin{acknowledgments}
This work was supported by NRF (National Honor Scientist Program 2010-0020414) and KISTI (KSC-2017-C3-0081). We thank D. C. Yang for reading the manuscript. 
\end{acknowledgments}


\bibliography{MPS_space_reduction}

\end{document}